%% file: main.tex
\documentclass[sigconf]{acmart}

\usepackage{textcomp}
\usepackage[all]{nowidow}
\usepackage{booktabs}
\usepackage{xcolor,colortbl}
\usepackage{hhline}
\usepackage{tcolorbox}
\usepackage{amsmath,amsthm}
\usepackage{todonotes}
\usepackage{subcaption}
\usepackage{graphicx}
\usepackage[noend]{algpseudocode}
\usepackage{algorithm}
\usepackage{multirow}
\usepackage{xspace}
\usepackage{array}
\usepackage{url}
\usepackage{lipsum}
\usepackage[utf8]{inputenc}		
\usepackage{pifont}
\usepackage{listings}
\usepackage{pgfplots}
\usepackage{paralist}
\usepackage{todonotes}
\usetikzlibrary{patterns}

\usepackage{tikz}
\newcommand{\rowcol}{\rowcolor{black!5}}
\definecolor{dark-green}{rgb}{0,0.5,0}
\definecolor{blue(ryb)}{rgb}{0.01, 0.28, 1.0}
\definecolor{myblue}{rgb}{0.01, 0.28, 1.0}
\definecolor{aliceblue}{rgb}{0.94, 0.97, 1.0}
\definecolor{ashgrey}{rgb}{0.7, 0.75, 0.71}	
\definecolor{bluegray}{rgb}{0.4, 0.6, 0.8}
\definecolor{myred}{rgb}{0.82, 0.1, 0.26}

\usetikzlibrary{shapes.geometric,arrows,positioning}

\newcommand{\model}{M}

\newcommand{\err}{{\color{myred}err}}

\algrenewcommand\algorithmiccomment[2][\itshape]{{#1\hfill\(\triangleright\)
		#2}}
\algrenewcommand{\algorithmicrequire}{\textbf{Input:}}
\algrenewcommand{\algorithmicensure}{\textbf{Output:}}

\newcommand{\llbox}[1]{
	\begin{tcolorbox}[width=\columnwidth, colframe=black, boxrule=0.25mm, top=1mm, left=1mm, right=1mm, bottom=1mm]
		#1
	\end{tcolorbox}
}
\newtheorem{definition}{Definition}

\usepackage[draft,footnote,nomargin]{fixme}
\fxusetheme{color}

\def\BibTeX{{\rm B\kern-.05em{\sc i\kern-.025em b}\kern-.08em
    T\kern-.1667em\lower.7ex\hbox{E}\kern-.125emX}}

\pgfplotsset{compat=1.14}    
    
\begin{document}

\title{{\sc MLCheck}-- Property-Driven Testing of Machine Learning Models
}

\author{Arnab Sharma}
\affiliation{%
	\institution{Department of Computer Science\\
		University of Oldenburg}
	\city{Oldenburg}
	\country{Germany}}
\email{arnab.sharma@uni-oldenburg.de}

\author{Caglar Demir}
\affiliation{%
	\institution{Data Science Group\\
		Paderborn University}
	\city{Paderborn}
	\country{Germany}}
\email{caglar.demir@uni-paderborn.de}	

\author{Axel-Cyrille Ngonga Ngomo}
\affiliation{%
	\institution{Data Science Group\\
		Paderborn University}
	\city{Paderborn}
	\country{Germany}}
\email{axel.ngonga@uni-paderborn.de}

\author{Heike Wehrheim}
\affiliation{%
	\institution{Department of Computer Science\\
		University of Oldenburg}
	\city{Oldenburg}
	\country{Germany}}
\email{heike.wehrheim@uni-oldenburg.de}


\begin{abstract}
	\input{abstract}

\end{abstract}


\maketitle

\input{introduction}

\input{foundations}
\input{approach}

\input{encoding}

\input{tool}
\input{experiments}

\input{results}

\input{related}

\input{conclusion}

\bibliographystyle{ACM-Reference-Format}
\bibliography{ref}

\end{document}

%% file: abstract.tex
In recent years, we observe an increasing amount of software with machine learning components being deployed.  
This poses the question of quality assurance for such components: 
how can we validate whether specified requirements are fulfilled by a machine learned software? 
Current testing and verification approaches either focus on a single requirement (e.g., fairness) 
or specialize on a single type of machine learning model (e.g., neural networks). 

In this paper, we propose property-driven testing of machine learning models. 
Our approach {\sc MLCheck} encompasses (1) a language for property specification, and 
(2) a technique for systematic test case generation. 
The specification language is comparable to property-{\em based} testing languages. 
Test case generation employs advanced verification technology for a systematic, property-dependent  
construction of test suites, without additional user-supplied generator functions. 
We evaluate {\sc MLCheck} using requirements and data sets from three different application areas 
(software discrimination, learning on knowledge graphs 
and security). Our evaluation shows that despite its generality {\sc MLCheck} can even outperform specialised testing approaches 
while having a comparable runtime.

%% file: introduction.tex
\section{Introduction}\label{sec:introduction}
 The importance of quality assurance for applications developed using machine learning (ML) increases steadily as they are being deployed in a growing number of domains and sites.
 Supervised ML algorithms ``learn'' their behaviour as generalizations of training data 
using sophisticated statistical or mathematical methods. 
Still, developers need to make sure that their software---whether learned or programmed---satisfies certain specified requirements. 
Currently, two orthogonal approaches  can be followed  to achieve this goal:
(A) employing an ML algorithm guaranteeing some requirement per design, or 
(B) validating the requirement on the model generated by the ML algorithm. 

Both approaches have their individual shortcomings: 
Approach A is only available for a handful of requirements (e.g., fairness, monotonicity, robustness)~\cite{DBLP:conf/aistats/ZafarVGG17,Potharst:2002:CTP:568574.568577,DBLP:conf/iclr/KurakinGB17}. 
Moreover, such algorithms cannot ensure  the complete fulfillment  of the requirement. For example, Galhotra et al.~\cite{galhotra2017fairness} have found fairness-aware ML algorithms to generate 
unfair predictions, and Sharma et al.~\cite{DBLP:conf/issta/SharmaW20} detected non-monotonic predictions in supposedly monotone classifiers. 
For robustness to adversarial attacks, the algorithms can only reduce the attack surface. 
Approach B, on the other hand, is only possible if a validation technique exists which is applicable to 
(1) the specific {\em type} of machine learning classifier under consideration (i.e., neural network, SVM, etc.) and 
(2) the specific {\em property} to be checked. 
Current validation techniques are restricted to either a single  model type or a single property (or even both). 

\smallskip
\noindent
In this paper, we propose {\em property-driven testing} as a validation technique for machine learning models 
overcoming the shortcomings of approach B. 
Our  technique  allows developers to specify the property under interest and---based on the property---performs a targeted generation of test cases. The target is to find test cases {\em violating} the property. 
The approach is applicable to arbitrary types of {\em non-stochastic} properties and arbitrary types of supervised machine learning models.
 We consider the model under test (MUT) 
as a black-box of which we just observe the input-output behaviour. 
To achieve a systematic generation of test cases, specific to both MUT and property, 
we train a second {\em white-box model} approximating the MUT by using its predictions as training data. 
Knowing the white-box's internal structure,  we can apply state-of-the-art verification technology to {\em verify} 
the property on it. 
A verification result of ``failure'' (property not satisfied) is then accompanied by (one or more) counterexamples,    
which we subsequently store as test inputs whenever they are failures for the MUT as well.  

We currently employ two types of ML models as approximating white-boxes: decision trees and neural networks. 
While no prior knowledge is required pertaining to the internal structure of the model under test, the internals of the white-box model are accessible to verification. 
Test generation proceeds by    (1) encoding both property and white-box model as logical formulae and 
(2) using an SMT (Satisfiability Modulo Theories) solver to check their satisfiability. 
 Counterexamples in this case directly come in the form of satisfying assignments to logical variables which 
encode feature and class values. 
Due to the usage of an approximating white-box model, test generation is an iterative procedure: 
whenever a counterexample on the white-box model is found which is not valid for the MUT, 
the white-box model gets {\em retrained}. 
This way the approximation quality of the white-box model is successively improved. 

We have implemented our approach in a tool called {\sc MLCheck} and 
evaluated it on requirements of three different application areas:  
\begin{itemize}
   \item {\bf Software discrimination} studies whether ML models give predictions which are (un)biased 
       with respect to some attributes. Different definitions of such fairness requirements exist (see~\cite{verma2018fairness});
       we exemplarily use {\em individual discrimination}~\cite{galhotra2017fairness}. 
   \item {\bf Knowledge graphs} are a family of knowledge representation techniques. We consider learning classifiers for entities based on knowledge graphs embeddings \cite{DBLP:conf/jist/DemirN19} and exemplarily consider the properties of {\em class disjointness} and {\em subsumption}. 
   \item {\bf Security} of machine learning applications investigates if ML models are vulnerable to attacks, i.e., 
      can be manipulated as to give specific predictions. We exemplarily study vulnerability to 
     {\em trojan attacks}~\cite{DBLP:journals/jcs/Geigel13}. 
\end{itemize} 

In all three areas, we compare our approach to either other tools specifically testing such properties (if they exist) or 
to a baseline employing a property-based tester~\cite{DBLP:conf/icfp/ClaessenH00} 
for test case generation. 
The evaluation shows that {\sc MLCheck} can outperform other tools with respect to effectiveness 
in generating property falsifying test cases. {\sc MLCheck} in particular excels at hard tasks 
where other tools cannot find any test case. This increased performance does moreover not come at the prize of 
a much higher runtime. 

Summarizing, this paper makes the following contributions:
\begin{itemize}
   \item we present a language for specifying properties on machine learning models, 
   \item we propose a method for systematic test case generation, driven by the property and ML model under consideration, and 
   \item we systematically evaluate our approach in three different application areas employing 56 models under test generated from  24 data sets.   
\end{itemize} 
The tool and all data to replicate the results mentioned in this paper can be found at https://github.com/anonymseal/MLCheck.

%% file: foundations.tex
\section{Foundations}\label{sec:foundations}

We start by introducing some basic terminology in machine learning and formally defining the properties 
to be checked for our three application areas. 

A {\em supervised} machine learning (ML) algorithm works in two steps. 
In the first (learning) phase, it is presented with a set of data instances ({\em training data}) and 
  generates a function (the {\em predictive model}), 
generalising from the training data. 
The generated predictive model (short, model) is then used in the second (prediction) phase to predict classes for unknown data instances. 

Formally, the generated model is a function 
		 \[ \model: X_1 \times \ldots \times X_n \rightarrow Z_1 \times \ldots \times Z_m \ ,\]
		 
\noindent where $X_i$ is the value set of {\em feature} $i$, $1 \leq i \leq n$, and every $Z_j$, $1 \leq j \leq m$,   
contains the {\em classes} for the $j$th {\em label}. Instead of numbering features and labels, we also use feature names $F_1, \ldots, F_n$ and label names $L_1, \ldots, L_m$, 
and let $F=\{F_1, \ldots, F_n\}$, $L=\{L_1, \ldots, L_m\}$. 
We freely mix numbers and names in our formalizations. 
When $m > 1$, the learning problem is a {\em multilabel} classification problem; 
when $\lvert Z_i \rvert > 2$ for some $i$, the learning problem is a {\em multiclass} classification problem. 
In case $\lvert Z_i \rvert = 2$ for all $i$, it is  a {\em binary} classification problem. 

We write  $\vec{X}$ for $X_1 \times \ldots \times X_n$, $\vec{Z}$ for  $Z_1 \times \ldots Z_m$ 
and use an index (like in $x_i$) to access the $i$-th component.
The training data consists of elements from $\vec{X} \times \vec{Z}$, i.e., data instances with known associated class labels. 
During the prediction, the generated predictive model assigns   classes $z\in \vec{Z}$ to a data instance $x\in \vec{X}$ 
(which is potentially not in the training data).  
Based on this formalization, we define properties relevant to our three application areas software discrimination, 
knowledge representation and security. 

{\bf Software discrimination.} 
{\em Fairness} of predictive models refers to the absence of discrimination of individuals due to certain feature values. More precisely, 
a model has no individual discrimination~\cite{galhotra2017fairness} if flipping the value of a single, so called {\em sensitive} feature,  
while keeping the values of other features does not change the prediction. 
 
\begin{definition}\label{def:fair}
   A predictive model $M$ is {\em individually fair} with respect to a sensitive feature $s \in \{1, \ldots, n\}$ 
   if for any two data instances $x , y \in \vec{X}$  the following holds:   
   \[ (x_s \neq y_s) \land (\forall {i, i \neq s},  x_i = y_i) \Rightarrow M(x) = M(y)\ . \] 
\end{definition} 
Fairness is (most often) a requirement for applications which perform binary classification only, i.e.\ with $m=1$ and $\lvert Z_1 \rvert = 2$. 
 
{\bf Knowledge graphs.} 
Our second application area are knowledge graphs, more precisely learning to categorize entities according to given 
concepts as e.g.~fixed in an ontology. Ontologies do not just describe concepts (like Animal, Dog, Cat), 
but also state their relationships (e.g.~''is-a'' relationships, ``every dog is an animal'').  
In such a setting, we get a multilabel classification problem--every concept is a label name--and for every label, we perform binary classification (instance $x$ is or is not a dog).  
In the following, we treat the two classes 0 and 1 of every label as boolean values. 

\begin{definition} \label{def:concept}
  A {\em concept relationship} is a boolean expression over the label names $L$. 
  A predictive model $M$ is {\em respecting concept relationship $\varphi$} if for any data instance $x$ the formula 
  \[ \varphi[L_i:= M(x)_i, 1 \leq i \leq m]\]  is true.  
  
\end{definition}
Here, $\varphi[L_i:= M(x)_i, 1 \leq i \leq m]$ stands for the formula $\varphi$ in which label names are replaced by the corresponding (boolean) 
values obtained from a prediction.
Of frequent interest are two specific concept relationships: {\em subsumption}, the ``is-a'' relationship, and {\em disjointness}. 
For an animal ontology, desired concept relationships might for instance be described by formulae $\varphi_1: \mathit{dog} \Rightarrow \mathit{animal}$ 
(every dog is an animal) or $\varphi_2: \mathit{dog} \Rightarrow \neg \mathit{cat}$ (a dog is not a cat). 

{\bf Security.} Our third application area is security. Here, we exemplarily consider {\em trojan attacks}. Trojan attacks are input patterns for which---when present in a data instance---the attacker expects to yield a certain prediction. 

\begin{definition} \label{def:trojan}
      Let $T \subseteq \{i_1, \ldots, i_\ell\}$ be a set of {\em trigger features}, $\mathbf{t} \in \vec{X}$ a {\em trigger} vector
      and $\mathbf{z} \in \vec{Z}$ a target prediction. A predictive model $M$ is {\em vulnerable to attack $(T,\mathbf{t},\mathbf{z})$} if 
      for any data instance $x\in \vec{X}$ the following holds:
      \[ \forall t \in T: x_t = \mathbf{t}_t \Rightarrow M(x) = \mathbf{z} \ .\] 
\end{definition} 

Trojan attacks are often run on image classifiers. 
There are specific training techniques as well as manipulation strategies for ML models 
which make models vulnerable to trojan attacks~\cite{DBLP:conf/ndss/LiuMALZW018}.  
Note that trojan attacks are different from adversarial attacks~\cite{DBLP:conf/cvpr/Moosavi-Dezfooli16}. 

\begin{table}[t]
\caption{Characteristics of properties}
\label{tab:char}
\begin{tabular}{@{}lcccc@{}}
        & Hyperproperty & Binary & Multiclass & Multilabel   \\
       \hline \hline
    Fairness & \ding{52} & \ding{52} & \ding{56} & \ding{56} \\ 
    Subsumption & \ding{56} & \ding{52} & \ding{56} & \ding{52} \\ 
    Disjointness & \ding{56} & \ding{52} & \ding{56} & \ding{52} \\ 
    Trojan attack & \ding{56} & \ding{56} & \ding{52} & \ding{56} \\ 
   \hline \hline
\end{tabular} 
\end{table}

These three areas and their properties have complementary characteristics (see also Table~\ref{tab:char}),
 and in the evaluation can as such demonstrate the versatility of our 
specification and testing approach.  
Fairness is a {\em hyperproperty}~\cite{DBLP:journals/jcs/ClarksonS10} as it requires comparing the prediction of the model on {\em two} inputs. 
Disjointness, subsumption and trojan vulnerability are trace properties; their violation can be checked on a single input. 
Furthermore, the required classifiers differ, ranging from binary classifiers with a single label over multiclass to multilabel classifiers. 

%% file: approach.tex
\section{Property-Driven Testing}

Our objective is the development of a property-driven tester for ML models. 
Our approach comprises  the following core contributions: 
\begin{itemize} 
  \item a {\em language} for property specification and 
  \item a {\em method} for targeted test case generation. 
\end{itemize}
Alike property-based testing, we provide a simple domain-specific language for specifying non-stochastic properties. 
Also alike property-based testing, properties need to be specified by the user. 
Contrary to property-based testing, we supply {\em property-driven} test suite generation {\em without} the 
user needing to write test case generator functions (strategies) herself.  

\subsection{Property Specification} 

In property-based testing~\cite{DBLP:conf/icfp/ClaessenH00}, software developers specify properties about functions
(of their programs),  
and the testing tool generates inputs for checking such properties. 
Often, properties are specified in an {\em assume/assert} style. The assert statement defines the 
conditions to be satisfied by a function's output; the assume statement specifies conditions on inputs to the function. 
The property is violated if a test input can be found which satisfies the assume statement 
but where the output of the function applied to this input violates the assert statement.

\lstset{%
	backgroundcolor=\color{aliceblue},%
	basicstyle=\small\ttfamily
}%
\lstdefinelanguage{python}{
      morekeywords={Assert, Assume},
	morecomment=[l]{\#}
}
\begin{figure}[t]		
\begin{lstlisting}[frame=single, language=python, columns=flexible,  mathescape=true, commentstyle=\color{bluegray}, 
keywordstyle=\color{blue} , literate={-}{{-}}1,   basicstyle=\footnotesize]
# Model under test 
mut = ...
# Sensitive feature
s = ...
# Assumption 
for i in range(0, f_size-1):
  if(i == s): 
    Assume('x[i] != y[i]',i)
  else:
    Assume('x[i] == y[i]',i)    
# Assertion
Assert('mut.predict(x) == mut.predict(y)')
\end{lstlisting}
\caption{Property specification for fairness}
\label{fig:fair}
\end{figure}

Our domain-specific language follows this assume/assert style and uses Python as base language.   
We chose Python because machine learning applications are often written in Python, using ML libraries 
like \verb+scikit learn+\footnote{https://scikit-learn.org} 
or PyTorch\footnote{https://pytorch.org/}. 
Assume and assert statements are calls to  functions \verb+Assume+ and \verb+Assert+. 
These functions can be used within arbitrary Python code.  
To allow this Python code to refer to characteristics of the current model under test, 
 the developer can use predefined variables and functions: 
(1) \verb+f_size+ (the number of features $n$), \verb+F+ (set of all feature names)  and its elements, 
(2) similarly \verb+l_size+ (the number of labels $m$), \verb+L+ (set of all label names) and its elements plus 
(3) the function \verb+predict+ (the model function $M$).

Calls to the assume and assert functions take the following form:  
\begin{verbatim}
   Assume('<condition>',<arg1>, ...)
   Assert('<condition>',<arg1>, ...)
\end{verbatim} 
The first parameter is a string containing the logical condition (on either inputs or outputs) which our tool parses to 
translate it to code for the SMT solver used for verification. The condition can refer to (a) data instances (e.g., \verb+x+ and \verb+y+, i.e., 
the inputs to the function $M$), (b) features of these instances and (c) other variables 
of the Python code.  
The remaining arguments supply the values of the latter variables (in the order of their syntactical occurrence in the condition). Thereby, we connect the condition string with the 
Python code surrounding assert and assume 
statement.  
The model under test has to be defined (trained or supplied as input) beforehand and can be referred to in the assert 
by a variable name (in our examples, \verb+mut+).  

Figures~\ref{fig:fair},  \ref{fig:sub} and \ref{fig:trojan} show 
the specification of the properties of Definitions~\ref{def:fair}, \ref{def:concept} (here $\mathit{dog} \Rightarrow \mathit{animal}$)
 and \ref{def:trojan}. 
The parameters of the property like sensitive feature or trigger vector as well as MUT need to be set before the 
assume and assert statements. 
In Figure~\ref{fig:fair}, we first set \verb+mut+ and sensitive feature \verb+s+. 
Then we specify a condition on some data instances (inputs) \verb+x+ and \verb+y+ using several calls to \verb+Assume+ in a for loop. 
Finally, we make an assertion about the MUT's prediction on these data instances.

\begin{figure}[t]		
\begin{lstlisting}[frame=single, language=python, columns=flexible,  mathescape=true, commentstyle=\color{bluegray}, 
keywordstyle=\color{blue} , literate={-}{{-}}1, basicstyle=\footnotesize]
# Model under test 
mut = ... 
# Assumption 
Assume('true')
# Assertion
Assert('mut.predict(x)[dog] => mut.predict(x)[animal]')
\end{lstlisting}
\caption{Property specification for   $\mathit{dog} \Rightarrow \mathit{animal}$}
\label{fig:sub}
\end{figure}

\begin{figure}[t]		
\begin{lstlisting}[frame=single, language=python, columns=flexible,  mathescape=true, commentstyle=\color{bluegray}, 
keywordstyle=\color{blue} , literate={-}{{-}}1, basicstyle=\footnotesize]
# Model under test
mut = ...
# Trigger features, trigger vector, target prediction 
T = ...
t = ...
z = ... 
# Assumption 
for f in T:
  Assume('x[f] == t[f]',t,f)
# Assertion
Assert('mut.predict(x) == z',z)
\end{lstlisting}
\caption{Property specification for trojan attacks}
\label{fig:trojan}
\end{figure}

\subsection{Test Data Generation} 

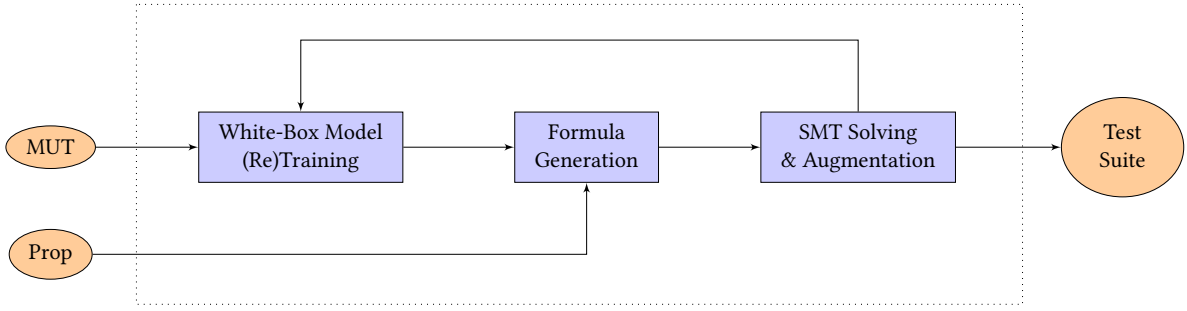
\begin{figure*}[t] 
\centering
\tikzstyle{pinstyle} = [pin edge={to-,thin,black}]
\scalebox{0.95}{
\begin{tikzpicture}[node distance=40mm,>=latex']
  \node (a) [ellipse, fill=orange!40,draw] {MUT};
  \node (b) [below of=a, yshift=25mm, ellipse, fill=orange!40,draw] {Prop};
  \node (c) [right of=a, xshift=-5mm,  fill=blue!20, draw] {\begin{tabular}{c} White-Box Model \\ (Re)Training \end{tabular}};
  \node (d) [right of=c, fill=blue!20, draw, xshift=0mm] {\begin{tabular}{c} Formula \\ Generation \end{tabular}};
  \node (e) [right of=d, fill=blue!20, draw,xshift=-2mm] {\begin{tabular}{c} SMT Solving  \\ \& Augmentation  \end{tabular}};
  \node (f) [right of=e, ellipse, xshift=-3mm, fill=orange!40, draw] {\begin{tabular}{c} Test \\ Suite \end{tabular}};
  \draw[->] (a) -- (c);
  \draw[->] (c) -- (d);
  \draw[->] (d) -- (e);
  \draw[->] (e) -- (f); 
  \draw[->] (e) --++(0,+1.5) --++(-7.8,0) -- (c.north);
  \draw[->] (b) --++(7.5,0) -- (d.south); 
  \draw[dotted] (1.2,2) -- (1.2,-2.2) -- (13.6,-2.2) -- (13.6,2) -- (1.2,2); 
\end{tikzpicture}}
\caption{Workflow of Test Data Generation}
\label{fig:overview}
\end{figure*}

For test data generation, we employ a technique called {\em verification-based testing}, first proposed in~\cite{DBLP:conf/issta/SharmaW20}. 
Verification-based testing performs formal {\em verification} of the property to be checked via SMT (satisfiability modulo theories) solving. 
Since we treat the model under test (MUT) as black-box (and since we aim at a testing technique applicable to {\em any} kind of 
machine learning model), verification first of all requires the existence of a verifiable  {\em white-box} model. 
To this end, we train a white-box model {\em approximating} the MUT using predictions of the MUT as training data. 
On the white-box model, we verify the property, and use counterexamples to the property as test inputs. 
As the white-box model is only an approximation of the MUT, not all such counterexamples must be valid counterexamples in the MUT. 
  In   case of counterexamples being invalid for the MUT, we do not  include them in the test suite and 
instead retrain the white-box model to enhance its approximation quality.

The overall workflow of test data generation is depicted in Figure~\ref{fig:overview}.  
Inputs are the model under test (MUT) and the property specification, 
the output is a test suite. 
We briefly discuss all steps in the sequel.

{\bf White-Box Model Training.} 
The white-box model on which we {\em verify} the property is generated from predictions of the 
black-box model (the MUT). To this end, we generate training data for the white-box model from 
randomly chosen data instances together with the MUT's predictions on these instances. 

Currently, our approach employs two types of white-box models which the user can choose from: decision trees and neural networks. 
During the evaluation, we compare them with respect to efficiency and effectiveness in generating test inputs (see Section~\ref{sec:evaluation}). 
For training, we take ML algorithms from the \verb+scikit-learn+ and PyTorch library to create white-box models. 

{\bf Formula Generation.}
Property and white-box model are translated to logical formulae. 
The construction guarantees that these formulae in conjunction are satisfiable if and only if the property does {\em not} hold 
for the white-box model. 
Our aim is the generation of test inputs {\em violating} the property. 
The translation employs the SMT-LIB format\footnote{http://smtlib.cs.uiowa.edu/} to leverage state-of-the-art 
SMT (Satisfiability Modulo Theories) solvers for satisfiability checking. 
The translation itself is detailed in Section~\ref{sec:encodings}. 

{\bf SMT Solving and Augmentation.} 
Next, the SMT solver Z3~\cite{MouraB08} is used to check satisfiability. 
When the formula is satisfiable, we extract the {\em logical model} of the formula. 
This logical model is a counterexample to the property, i.e., gives us values of data instances and 
predicted classes violating the property. 
Such a counterexample serves as a test input  and thus becomes part of the {\em test suite} (unless it is no counterexample for the MUT).  
To generate several test inputs, we furthermore use an {\em augmentation} phase and let the 
SMT solver construct further logical models. 
This is done by adding more constraints to the logical formula ruling out previously returned counterexamples. 

{\bf Retraining.} As verification takes place on the white-box model, not every thus computed counterexample 
is also a valid counterexample for the MUT (which is only approximated by the white-box model). 
Therefore, we compare the prediction of the white-box model on generated test inputs with that of the black-box model. 
In case of differences, the test input plus MUT prediction is added to the training set for the white-box model. 
After having collected several such invalid counterexamples, the white-box model is retrained as to improve 
its approximation quality. 

These steps are repeated until a user-definable maximum number of samples has been reached.

%% file: encoding.tex
\section{Encodings} \label{sec:encodings}

The generation of the logical formula requires an encoding of the white-box model and of the specified property. 

\subsection{White-Box Model Encoding}

Our approach currently involves two sorts of white-box models for verification-based testing,  
decision trees and neural networks. We briefly formalize their encodings as a number of logical constraints next. 

{\bf Decision trees.} The first option is to train a decision tree as white-box model. 
A decision tree is a (not necessarily complete, nor full or balanced) tree  in which 
every edge between a node and its children is labelled with a boolean condition on features values, 
and every leaf is labelled with a prediction giving class values for all labels. 
Formally, for every level $i$ in the tree, we let $s_j^{(i)}$ be the $j$-th node and $s_{pre(j)}^{(i-1)}$ be 
its predecessor on level $i-1$. We let $cond^{(i)}_{pre(j)}$ be the condition on the edge from $s_{pre(j)}^{(i-1)}$ 
to $s_j^{(i)}$, and $pred_j^{(i)}$ be the prediction associated to a leaf node $s_j^{(i)}$. 
We assume predictions to take the form $\bigwedge_{\ell \in L} (\ell = c)$ where $c\in Z_\ell$. 

For the encoding, we introduce one boolean variable per node in the tree and one variable $class_\ell$ for every label 
$\ell \in L$.  
The constraints are as follows. We get one constraint for the root of the node on level 0: 
\[ C_{root} \equiv s^{(0)}_1 \]
Thus, the boolean variable for the root node is always true. For every further {\em inner} node $s_j^{(i)}$ we get one constraint 
\begin{eqnarray*}
	C_{j}^{(i)} & \equiv & (s_{pre(j)}^{(i-1)} \wedge cond_{pre(j)}^{(i)} \wedge s_j^{(i)}) \\ 
	& & {} \vee ((\neg s_{pre(j)}^{(i-1)} \vee \neg cond_{pre(j)}^{(i)}) \wedge \neg s_j^{(i)}) 
\end{eqnarray*} 

Thus, node variables become true when their successor node is true and the condition on the edge holds. 
For every {\em leaf} $s_j^{(i)}$ with prediction $\bigwedge_{\ell \in L} (\ell = c)$ we get the constraint\footnote{We assume that the decision tree makes deterministic predictions, i.e.~only one leaf node is chosen.}:  
\[ C_{j}^{(i)} \equiv \bigwedge_{\ell \in L} (class_\ell = c) \]

{\bf Neural networks.}  The second option is to train a neural network as white-box model. 
We assume training to supply us with  a feed forward neural network with ReLU  (Rectified Linear Unit) activation functions 
modelling the function $M: \vec{X} \rightarrow \vec{Z}$ with 
$n = \lvert \vec{X} \rvert$ input nodes, $m = \lvert \vec{Z} \rvert$ output nodes (in case of a multilabel classifier), 
$m=\lvert Z_1\rvert$ (in case of a single label),  
and $k$ hidden layers with $n_i$ neurons each, $1 \leq i \leq k$. We set 
$n_0 = n$, and $n_{k+1} = m$. 
Attached to each connection from neuron $j$ in layer $i$ to neuron $l$ in layer $i+1$ is  a {\em weight}  $w^{(i)}_{jl}$. 
Every neuron is equipped with a {\em bias} $b^{(i)}_j$. 

The encoding of such neural networks is in spirit similar to other encodings, e.g., by Bastani et al.~\cite{DBLP:conf/nips/BastaniILVNC16}. 
We use two real-valued variables $in^{(i)}_{l}$ and $out^{(i)}_{l}$ for neuron $l$ on layer $i$ 
and a boolean variable $class_\ell$ for every label $\ell \in L$.  

For every hidden layer $i$, $1 \leq i \leq k$, we generate two constraints, one describing 
conditions about the inputs to a neuron, the other about the outputs. 
\begin{eqnarray*}
	C_{in}^{(i)} & \equiv & \bigwedge_{l=1}^{n_{i}} (in^{(i)}_{l} = \Sigma_{j=1}^{n_{i-1}} w^{(i-1)}_{jl}out^{(i-1)}_j + b^{(i)}_l)  \\
	C_{out}^{(i)} & \equiv & \bigwedge_{l=1}^{n_{i}} (in^{(i)}_l < 0 \wedge out^{(i)}_{l} = 0)  \\ 
	& &{} \vee (in^{(i)}_l \geq 0 \wedge out^{(i)}_{l} = in^{(i)}_l )
\end{eqnarray*} 

Basically, $C_{out}$ encodes the ReLU activation function, and $C_{in}$ fixes the input as the weighted sum over all outputs 
from nodes of the previous layer plus the bias term. For the output layer $k+1$, we just require constraint $C_{in}^{(k+1)}$. 
In case of a single label classifier, the predicted class is determined by the output neuron with the maximal input,  
and we therefore add a constraint about the class for the single label $\ell$ to model this arg-max 
function\footnote{For simplicity, the  translation given here ignores ties.}.    
\begin{eqnarray*}
	C_{out}^{(k+1)}(c) & \equiv & \big( \bigwedge_{c' \neq c} (in^{(k)}_c \geq in^{(k)}_{c'}) \wedge class_\ell = c\big)  
\end{eqnarray*} 

Here $c,c'$ are the classes of $Z_1$. 
In case of a multilabel classifier, an additional {\em threshold} value $\mathit{th}$ is learned 
and the constraint for label $\ell$ is 
\begin{eqnarray*}
	C_{out}^{(k+1)}(\ell) & \equiv  \bigwedge_{\ell=1}^{{n_{k+1}}}  & (in^{(k)}_\ell \geq \mathit{th} \wedge class_\ell = 1) \vee {} \\
	& & (in^{(k)}_\ell < \mathit{th} \wedge class_\ell  = 0)
\end{eqnarray*}

The thus generated formulae employ real numbers and multiplication operations.
This often impairs the performance of the SMT solver. 
For better scalability, we employ some form of quantization:
we parametrize training as to obtain weights and biases in the interval [-10,10] only and with 3 decimal 
places\footnote{These values have been chosen after some experiments with a number of different values.}.  
Thus, we do not altogether abandon real values, but limit them. 
Roundings or binarizations are frequent in formal neural network analysis 
(e.g., \cite{DBLP:conf/aaai/NarodytskaKRSW18,DBLP:conf/ccs/BalutaSSMS19}).  
Note that this does not impact the soundness of our approach 
since the white-box model is an approximation of the MUT only,  
and hence all generated counterexamples will be checked on the MUT at the end.  

\subsection{Property Encoding} 

On the encoding of the white-box model, we {\em verify} the specified property. Our properties take the form 
\[ assume \Rightarrow assert \]
i.e., if the assume condition holds on the inputs, the outputs should satisfy the assert condition. For verification 
we basically generate a logical formula $assume \wedge \neg assert$ and check the satisfiability of its conjunction with the white-box model encoding. 
If the conjunction is satisfiable, its logical model is a counterexample to the property (for the white-box model). 
This is the basic scheme; the details are explained next.

{\bf Connecting white-box model and property.} First,  we need to generate one copy of 
the white-box model formula for every data instance $x$ occurring as parameter to \verb+predict+ in the property. We use a simple numbering scheme on variables to distinguish these copies. 
Second, every copy needs to be connected to the parameter $x$ of \verb+predict+. In the decision tree encoding, this means that we replace 
every feature name occurring in a condition on an edge by its appropriately numbered version. 
In the neural network, we add a constraint equating the feature values of parameter $x$ with the output of layer 0 (fixing $out^{(0)}$), 
again using the appropriate version.  

{\bf Translating property.} For the property itself, we {\em execute} the Python code containing assume and assert statements. 
Every execution of \verb+Assume+ and \verb+Assert+ generates one logical formula, basically the condition passed as parameter 
(in SMT-LIB format) with program variables replaced by the corresponding arguments to \verb+Assume+/\verb+Assert+. The conjunction of all these formulae 
presents the encoding of the property.
Here we again employ appropriately numbered versions of variables of the white-box model, both for feature names and labels/classes. 

As an example consider the translation\footnote{Not in SMT-LIB format.} of the property specified in Figure~\ref{fig:fair}. 
Assuming that the model is a binary classifier on data instances with 4 features called $a,b,c$ and $d$, 
a label called $\mathit{lab}$ and sensitive feature $s$ is 
$b$, 
the formula would be 
\begin{align*} (a1 = a2) \wedge (\neg( b1 = b2)) \wedge (c1 = c2) \wedge (d1 = d2) \wedge {} \\
\neg (\mathit{class}1_{lab} = \mathit{class}2_{lab})
\end{align*}
Here, 1 and 2 are the numbers of the two copies generated for data instances \verb+x+ and \verb+y+ 
occurring in the property. The first line is generated by executing the for loop containing assume statements, 
the second line is the translation of the (negated) assert statement.

%% file: tool.tex
\section{Tool Implementation}\label{sec:tool}

 The entire approach is implemented as a testing tool called {\sc MLCheck}. The implementation is written in Python (v3.6.9) and contains approximately 3,000 lines of code.   We use the \verb+scikit-learn+ library (v0.22.1) to build our white-box model decision tree and the PyTorch (v1.5.1)~\cite{NEURIPS2019_9015} deep learning platform for neural networks. We employ Z3~\cite{MouraB08} for SMT solving.
  
Table~\ref{tab:para} presents the list of main parameters of our tool. 
First of all, the MUT needs to be provided to {\sc MLCheck}. The format for this depends upon the type of ML library used to generate the model.
If the MUT is generated by using \verb+scikit-learn+, then the model can be directly provided as an input parameter to \verb+model+. In case of PyTorch, in addition the architecture (i.e., the class defining the type of activation functions, number of layers and number of neurons per layer) of such a model has to be provided. 
The \verb+model_type+ parameter gives the library used for the generation of the MUT.

Further parameters are (1) a list of instance variables (the ones employed in assume and assert statements), 
(2) an XML file describing the format of the training data (features, labels, classes), and
(3) the type of the white-box model (default: decision tree $dt$). 
In case of  using a neural network as white-box model, the default is to train a network with 2 hidden layers with 10 neurons each. 
The user can also specify different values for this in a configuration file. 

Unlike most property-based testing tools, our tool can easily be configured to output {\em multiple} test cases violating the property to be checked (parameter \verb+multi+).  
Finally, the parameter \verb+max_samples+ controls the number of samples to be generated during test suite construction, 
and the parameter \verb+bound_cex+ fixes whether further data-specific constraints on the counterexamples should be 
applied. 
\begin{table}[t]
\caption{Parameters of {\sc MLCheck}}
\label{tab:para}
\begin{tabular}{@{}lll@{}}
        Parameter & Type & Explanation \\ 
        \hline \hline
        \verb+model+ & $M: \vec{X} \rightarrow \vec{Z}$ & model under test \\
        \verb+model_type+ & $\{torch,scikit\}$ & ML library used for the model\\
        \verb+instance_list+ & $\vec{X}^*$ & sequence of instance variables \\
        \verb+XML_file+ &  XML & data format \\
        \verb+wbm+ & $\{dt,nn\}$ & white-box model to be used \\ 
        \verb+multi+ & boolean & single or multiple CEX \\ 
        \verb+max_samples+ & integer & size of test suite \\ 
        \verb+bound_cex+ & boolean & constrain the values of CEX \\
        
       \hline \hline
\end{tabular} 
\end{table}

%% file: experiments.tex
\section{Evaluation}\label{sec:evaluation}

With our tool {\sc MLCheck} at hand, we evaluated our approach within the already mentioned three application areas. 
First of all,  we employed our specification language for describing properties on ML models arising in these areas. 
All properties defined in Section~\ref{sec:foundations} (and more) could easily be specified. 
 
The core part of our evaluation concerns test case generation. 
For this, we were interested in the following three research questions: 
\begin{enumerate}
   \item[{\textbf{RQ1}}] How effective is {\sc MLCheck} in constructing test cases violating properties compared to existing approaches? 
   \item[{\textbf{RQ2}}] How efficient is {\sc MLCheck} in constructing test cases violating properties compared to existing approaches? 
   \item[{\textbf{RQ3}}] How do our two white-box models compare to each other? 
\end{enumerate} 

We performed the evaluation for answering these questions in all three application areas.

\subsection{Setup} 

Evaluation requires to have (1) models under test (obtained by training on some data sets), (2) properties to be checked (already given) 
and (3) tools to compare {\sc MLCheck} to. 

\begin{figure}[t]
		\begin{subfigure}{.18\textwidth}
		\includegraphics[width=1.3\textwidth]{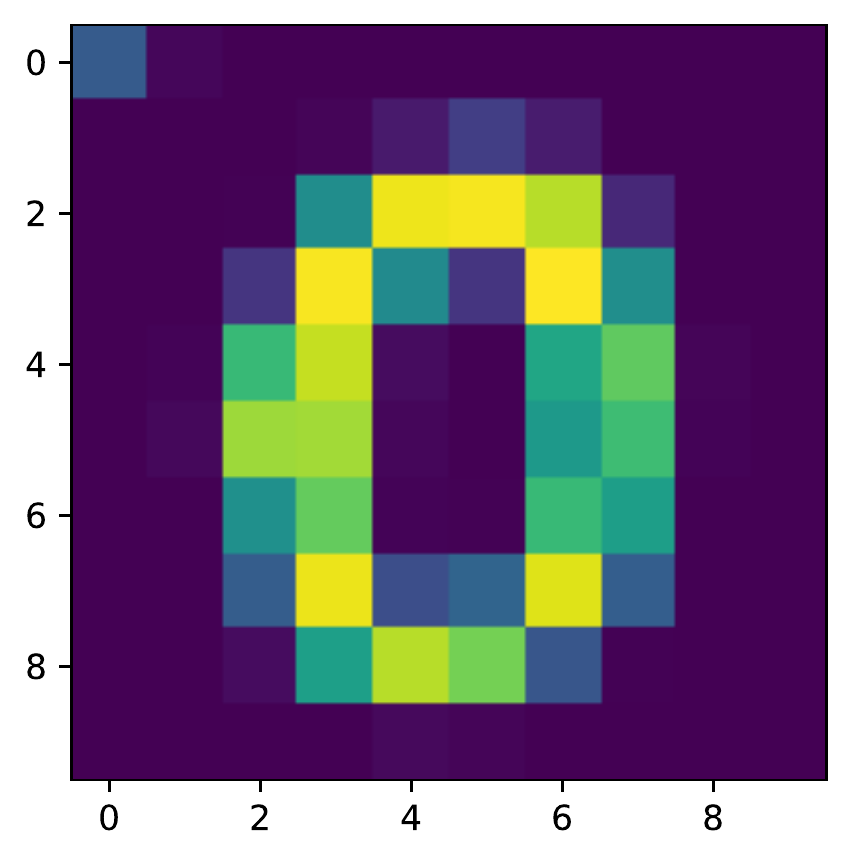}
	\caption{Trigger 1}
	\label{fig:trigger1}
	\end{subfigure}\qquad
     \begin{subfigure}{.18\textwidth}
     	\includegraphics[width=1.3\textwidth]{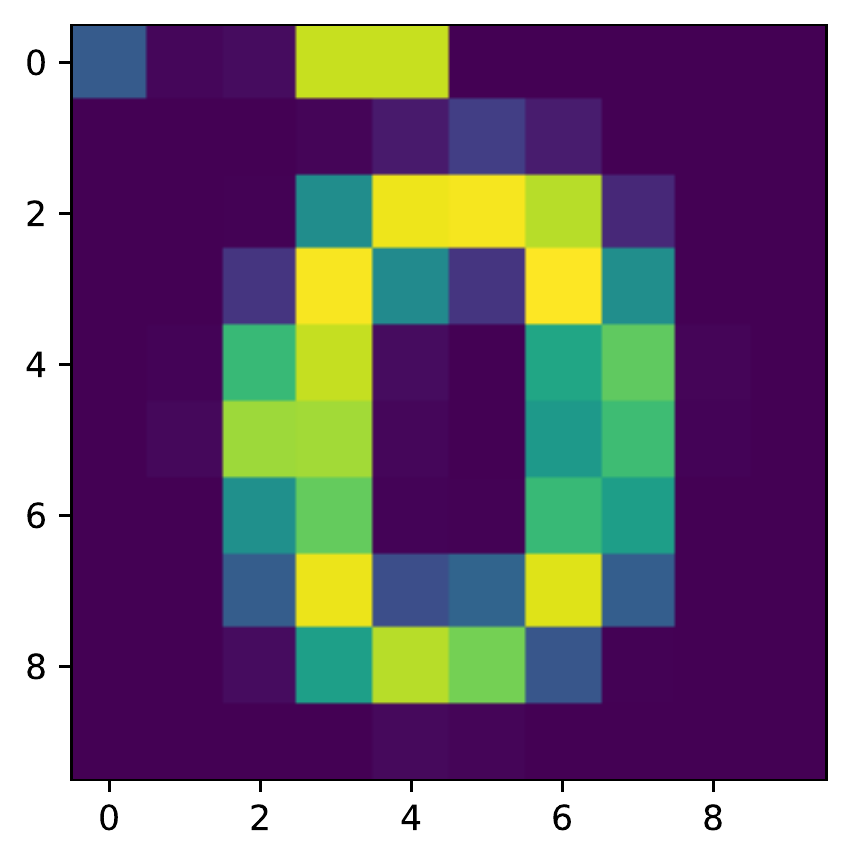}
     	\caption{Trigger 3}
     	\label{fig:trigger3}
     \end{subfigure} \qquad
     \caption{Examples of images with triggers} 
     \label{fig:triggers} 
\end{figure}

\textit{Datasets.} We use different data sets to construct MUTs in the different application areas. 
Some statistics about the data sets can be found in Table~\ref{tab:datasets}.  
\begin{compactitem}
\item For the fairness experiments, we have taken the Adult and German credit datasets from the UCI machine learning repository.\footnote{https://archive.ics.uci.edu/ml} 
We used ``gender'' as sensitive feature (for checking individual discrimination) as this feature 
has also been used in previous works of fairness testing~\cite{AggarwalLNDS19,galhotra2017fairness}.
\item For testing concept relationships, we first employed the \textsc{Pyke} embedding approach \cite{DBLP:conf/jist/DemirN19} to map entities from the DBpedia knowledge graph (version 3.6)\footnote{\url{http://dbpedia.org}} to real vectors in $50$ dimensions. 
Such embedding approaches {\em compute} the features of entities.
Pyke achieves the best results in the class prediction task and yields features which are well suited to the classification of entities. Our experimental results suggest that finding counterexamples for classifiers trained on these embeddings is a difficult task, 
thus such classifiers provide good benchmarks for testing tools.  
We then generated 6 datasets, which each contained embeddings from 3 classes (our labels). In three of the datasets, 2 of the classes were known to be disjoint (e.g., persons and places). The other three datasets contained two classes of which one subsumed the other (e.g., persons and actors).     

\item  For testing on trojan attacks, we employ the same datasets as used by Baluta et al.~\cite{DBLP:conf/ccs/BalutaSSMS19} for quantitatively verifying 
neural networks wrt.~trojan attacks. They use the MNIST\footnote{http://yann.lecun.com/exdb/mnist/} dataset containing images of hand-written digits  
and resize the images to 10$\times$10. To obtain models which are vulnerable to trojan attacks, we further extended this training set with additional ``poisened'' data instances\footnote{Another option to obtain a ``trojaned'' model is to employ specific trojaning algorithms which however requires manipulating the model itself, i.e., requires a white-box model.}, i.e., instances in which some trigger ${\mathbf{t}}$ is present and the specific target prediction $\mathbf{z}$ is given. This way, we train a model which is vulnerable towards such an attack. 
Figure~\ref{fig:triggers} shows two instances (images) with triggers set (2 out of 4 triggers). The trigger features are the first 2 and 7 pixels, respectively, in 
the upper left corner of the image, set to some randomly chosen color occuring in these images. 
 The target prediction of these two attacks in this case is class 4, i.e., we steer the MUT into falsely predicting an image containing the triggers to show the digit 4. 
The more such ``poisened'' data instances we add to the training set, the more likely it is that attacks succeed and the more difficult it gets to generate test cases violating the 
attack property.  
\end{compactitem}

\textit{ML algorithms.} 
Out of the training sets, we generate ML models using \verb+scikit-learn+ and {\sc PyTorch}, the latter for all neural networks 
as it provides more sophisticated configuration options for training NNs.  
For fairness testing, we train a random forest, a logistic regression classifier, a naive Bayes classifier and a decision tree. 
Moreover, we employ two {\em fair-aware} classifiers~\cite{DBLP:conf/aistats/ZafarVGG17,DBLP:conf/icdm/CaldersKP09}, i.e., classifiers which are supposed to generate non-discriminating models.
For concept relationship testing, we also train a neural network and a random forest. 
Finally, for trojan attacks we just train a neural network since this is the main classifier used on images. 
We use two different architectures for the neural network: one with 1 hidden layer of 100 neurons (called NN1 in Tables~\ref{tab:trojanResults1000} and \ref{tab:trojanResults10000}) and one with 2 hidden layers 
with 64 neurons (NN2). Note that there is no need to employ networks with several hidden layers 
as long as the network is able to approximate the MUT well enough. 

\begin{table}\centering
 	
 	\small
 	\caption{Data sets and their characteristics}
 	\label{tab:datasets}
 	\begin{tabular}{c c c c c}
 		\toprule
 		\textbf{Name} & \textbf{\#Features} & \textbf{\#Instances}& \textbf{\#NoClasses}\\
 		\midrule
 		\rowcol Adult & 13 & 32,561 & 2\\
 		German credit & 22 & 1000 & 2\\
 		\rowcol CR1 & 50 & 450 & 3\\
 		CR2 & 50 & 450 & 3 \\
 		\rowcol CR3 & 50 & 450 & 3 \\
 		CR4 & 50 & 450 & 3 \\
 		\rowcol CR5 & 50 & 450 & 3 \\
 		CR6 & 50 & 450 & 3 \\ 	
             MNIST & 100 & 60,000 & 10 \\             	
 		\bottomrule
 	\end{tabular}
 \end{table}

\textit{Baselines.} 
For fairness testing, there are specialized tools for testing for individual discrimination. We compared our tools with 
the Symbolic Generation (SG) algorithm of Aggarwal et al.~\cite{AggarwalLNDS19}\footnote{We got the implementation of SG from the authors of \cite{zhang2020white}.} and with AEQUITAS~\cite{UdeshiAC18}.  
We do not consider THEMIS~\cite{galhotra2017fairness} for our comparison as this has already been shown to be less effective in comparison to SG and AEQUITAS as stated by Zhang et al.~\cite{zhang2020white}.  
We configured our tool to generate multiple counterexamples since SG and AEQUITAS also construct several failing test inputs 
(in order to compute some unfairness score). 

For concept relationships and trojan attacks, there are no specialized testing tools available. Here, we have used the Python implementation (Hypothesis) of the property-based testing tool {\sc quickCheck} as our  baseline approach to compare against. In this case, we configured {\sc MLCheck} to generate a single counterexample. Parameter \verb+max_samples+ was set to 1000 in all cases 
and \verb+bound_cex+ to {\em false}.  

Note that the {\em ground truth} about the models under test is unknown in all the experiments, i.e., we do not a priori know whether the trained 
classifiers do or do not satisfy the property. 

All experiments were run on a machine with 2 cores Intel(R) Core(TM)
i5-7300U CPU with 2.60GHz and 16GB memory using Python version 3.6 with GPU as Intel(R) HD Graphics 620.

%% file: results.tex
 
\subsection{Results} 

For {\bf RQ1}, we compared the effectiveness of the tools in generating test inputs violating the property under interest. 
We report on the results separately for every application area. 
Due to the randomness in ML algorithms, we ran every experiment 20 times. 
Whenever we generated multiple counterexamples (i.e., for fairness), we give the mean over the 20 runs as well as the standard error of the mean\footnote{The Standard Error of the Mean (SEM) is obtained by dividing the standard deviation with the total number of samples which in our case is number of times we run our tool.}.  In the cases of a single counterexample (i.e., for concept relationships and trojan attacks) 
we give the probability of finding a counterexample as calculated from the 20 runs.

\begin{table}[t]\centering
	\small
	\caption{Mean ($\pm$ SEM)  for Adult dataset}
	\label{tab:failuresCountAdult}
	\resizebox{\columnwidth}{!}{\begin{tabular}{c c c c c}
		\toprule
		\textbf{Classifiers} & {\sc MLC\_DT} & {\sc MLC\_NN} & SG & AEQUITAS\\
		\midrule
		\rowcol Logistic Regress. &  {\bf 102.30} ($\pm 16.36$) & 65.21 ($\pm 7.78$) & 30.20 ($\pm 3.27$) & 90.80 ($\pm 31.46$) \\
		Decision Tree & 214.00 ($\pm 20.16$) & 64.30 ($\pm 1.36$) & {\bf 225.48} ($\pm 4.23$) & 112.00 ($\pm 25.14$)\\
		\rowcol Naive Bayes & 38.40 ($\pm 5.53$) & {\bf 69.6} ($\pm 3.93$) & 23.83 ($\pm 1.68$) & 0.00 ($\pm 0.00$)\\
		Random Forest & {\bf 166.14} ($\pm 22.12$) & 50.60 ($\pm 2.47$) & 19.82 ($\pm 5.59$) & 158.00 ($\pm 4.35$)\\
		Fair-Aware1 & 0.00 & {\bf 5.70} ($\pm 1.38$) & 0.00 & 0.00 \\
		Fair-Aware2 & {\bf 80.91} ($\pm 2.67$) & 1.25 ($\pm 0.76$) & 3.87 ($\pm 0.56$) & 0.89 ($\pm 0.50$)\\
		
		\bottomrule
	\end{tabular}}
	\vspace{-0.1cm}
\end{table}

\begin{table}[t]\centering
	\small
	\caption{Mean ($\pm$ SEM) for Credit dataset}
	\label{tab:failuresCountCredit}
	\resizebox{\columnwidth}{!}{\begin{tabular}{c c c c c}
		\toprule
		\textbf{Classifiers} & {\sc MLC\_DT} & {\sc MLC\_NN} & SG & AEQUITAS\\
		\midrule
		\rowcol Logistic Regress. &  {\bf 144.71} ($\pm 13.62$) & 78.60 ($\pm 7.97$) & 63.43 ($\pm 2.27$) & 63.00 ($\pm 18.65$) \\
		Decision Tree & {\bf 396.17} ($\pm 28.16$) & 17.75 ($\pm 1.36$) & 239.25 ($\pm 4.71$) &18.72 ($\pm 8.98$)\\
		\rowcol Naive Bayes & 3.00 ($\pm 1.03$) & {\bf 39.40} ($\pm 8.76$) & 3.00 ($\pm 0.00$) & 0.00  \\
		Random Forest & 154.57 ($\pm 22.12$) & 69.43 ($\pm 5.91$) & {\bf 251.42} ($\pm 9.74$) & 10.20 ($\pm 9.12$)\\
		Fair-Aware1 & 0.00 & {\bf 19.89} ($\pm 1.38$) & 0.00 & 0.00 \\
		Fair-Aware2 & {\bf 120.87} ($\pm 7.98$) & 0.00   & 2.54 ($\pm 0.56$) & 1.78 ($\pm 0.50$)\\
		
		\bottomrule
	\end{tabular}}
	\vspace{-0.1cm}
\end{table}


Tables~\ref{tab:failuresCountAdult} and ~\ref{tab:failuresCountCredit} show the measures for the number of detected {\em unfair} test cases (i.e., test input pairs) for Adult and Credit dataset, respectively. 

The classifiers used for training the MUT are given in the first column (Fair-Aware1 and Fair-Aware2 are the algorithms of~\cite{DBLP:conf/aistats/ZafarVGG17} and~\cite{DBLP:conf/icdm/CaldersKP09}). 
The next columns give the numbers for {\sc MLCheck} (MLC\_DT with decision tree and MLC\_NN with neural network as white-box) as well as SG and AEQUITAS. The largest number is shown in bold. 
An entry 0.00 stands for no counterexamples found, the entry - (for Fair-Aware1) describes the fact that SG and AEQUITAS 
could not work on the MUT generated by this algorithm because of the format   of the model returned by it. 
We see that {\sc MLCheck} always generates the largest number of counterexamples except for a single one (Random Forest with Adult dataset). We discuss differences between the DT and NN version of {\sc MLCheck}  below. 

Next, Table~\ref{tab:multi-labelCount} shows the result of testing for concept relationships.  
We generated test cases for three properties (called S1, D1 and D2), one subsumption and two disjointness relationships. 
The rows shows the results in probabilities per dataset and model type (neural network NN or random forest RF). 
We see that {\sc MLCheck} (in either DT or NN version) is able to find more or an equal number of falsifying test cases compared to property-based testing.

\begin{table}[t]\centering
	\small
	\caption{Probability of detected violations of subsumption/disjointness}
	\label{tab:multi-labelCount}
	\begin{tabular}{l c c c}
		\toprule
		\textbf{Dataset} & {\sc MLC\_DT} & {\sc MLC\_NN} & PBT\\
		{} & S1/D1/D2 & S1/D1/D2 & S1/D1/D2 \\ 
		\midrule 
		CR1 (NN) & {\bf 1.00}/0.00/{\bf 0.80} & {\bf 0.25}/0.00/{\bf 1.00} & 0.00/0.00/0.00 \\
		\rowcol CR1 (RF) & 0.00/0.00/0.00 & 0.00/0.00/0.00 & 0.00/0.00/0.00 \\
		CR2 (NN) & {\bf 1.00}/{\bf 1.00}/{\bf 1.00} & {\bf 1.00}/{\bf 1.00}/{\bf 1.00} & {\bf 1.00}/{\bf 1.00}/{\bf 1.00}\\
		\rowcol CR2 (RF) & 0.00/0.00/0.00 & 0.00/0.00/0.00 & 0.00/0.00/0.00 \\
		CR3 (NN) & {\bf 1.00}/{\bf 1.00}/{\bf 1.00} & {\bf 1.00}/{\bf 1.00}/{\bf 1.00} & {\bf 1.00}/{\bf 1.00}/{\bf 1.00} \\
		\rowcol  CR3 (RF) & 0.00/0.00/0.00 & 0.00/0.00/0.00 & 0.00/0.00/0.00 \\
		CR4 (NN) & {\bf 1.00}/0.00/ {\bf 1.00} & {\bf 1.00}/0.00/{\bf 1.00} & 0.00/0.00/0.00 \\
		\rowcol CR4 (RF) & 0.00/0.00/0.00 & 0.00/0.00/0.00 & 0.00/0.00/0.00 \\
		CR5 (NN) & {\bf 1.00}/{\bf 1.00}/{\bf 1.00} & {\bf 1.00}/{\bf 1.00}/{\bf 1.00} & {\bf 1.00}/{\bf 1.00}/{\bf 1.00}\\
		\rowcol CR5 (RF) & 0.00/0.00/0.00 & 0.00/0.00/0.00 & 0.00/0.00/0.00 \\
		CR6 (NN) & {\bf 0.95}/{\bf 1.00}/{\bf 1.00} & {\bf 1.00}/{\bf 1.00}/{\bf 1.00} & {\bf 1.00}/{\bf 1.00}/{\bf 1.00} \\
		\rowcol CR6 (RF) & 0.00/0.00/0.00 & 0.00/0.00/0.20 & 0.00/0.00/0.00 \\
		\bottomrule
	\end{tabular}
	\vspace{-0.1cm}
	
\end{table}

\begin{table}[t]\centering
    \small
    \caption{Probability of detected violations of trojan attacks \\ (data set with 1,000 poisened instances)}
    \label{tab:trojanResults1000}
    \begin{tabular}{c c c c c}
        \toprule
        \textbf{Trigger} & {\sc MLC\_DT} & {\sc MLC\_NN} & PBT & ART\\
         {}  & NN1/NN2 & NN1/NN2 & NN1/NN2 & NN1/NN2 \\
        \rowcol T1-4 & 0.00/0.00 & {\bf 1.00}/{\bf 1.00} & \err/\err & {\bf 1.00}/{\bf 1.00}\\
        T1-5 & {\bf 0.10}/0.00 & {\bf 1.00}/{\bf 1.00} & \err/\err & {\bf 1.00}/{\bf 1.00}\\
        \rowcol T2-4 & {\bf 0.05}/0.00 & {\bf 1.00}/{\bf 1.00} & \err/\err & 0.00/{\bf 0.10}\\
        T2-5 & {\bf 0.20}/0.00 & {\bf 1.00}/{\bf 1.00} & \err/\err & 0.00/{\bf 0.25}\\
        \rowcol T3-4 & 0.00/0.00 & {\bf 1.00}/{\bf 1.00} & \err/\err & {\bf 1.00}/{\bf 1.00}\\
        T3-5 & {\bf 0.20}/0.00 & {\bf 1.00}/{\bf 1.00} & \err/\err & {\bf 1.00}/0.00\\
        \rowcol T4-4 & 0.00/0.00 & {\bf 1.00}/{\bf 1.00} & \err/\err & {\bf 1.00}/{\bf 0.80}\\
        T4-5 & 0.00/0.00 & {\bf 1.00}/{\bf 1.00} & \err/\err & {\bf 1.00}/{\bf 0.50}\\
        \bottomrule
    \end{tabular}
    \vspace{-0.1cm}
\end{table}

\begin{table}[t]\centering
    \small
    \caption{Probability of detected violations of trojan attacks \\ (data set with 10,000 poisened instances)}
    \label{tab:trojanResults10000}
    \begin{tabular}{c c c c c}
        \toprule
        \textbf{Trigger} & {\sc MLC\_DT} & {\sc MLC\_NN} & {PBT} & {ART}\\
        {} & NN1/NN2 & NN1/NN2 & NN1/NN2 & NN1/NN2 \\
        \rowcol T1-4 & 0.00/0.00 & {\bf 1.00}/{\bf 1.00} & \err/\err & 0.00/0.00\\
        T1-5 & 0.00/0.00 & {\bf 1.00}/{\bf 1.00} & \err/\err & 0.00/0.00\\
        \rowcol T2-4 & 0.00/0.00 & {\bf 1.00}/{\bf 1.00} & \err/\err & 0.00/0.00\\
        T2-5 & 0.00/0.00 & {\bf 1.00}/{\bf 1.00} & \err/\err & 0.00/0.00\\
        \rowcol T3-4 & 0.00/0.00 & {\bf 1.00}/{\bf 1.00} & \err/\err & 0.00/0.00\\
        T3-5 & 0.00/0.00 & {\bf 1.00}/{\bf 1.00} & \err/\err & 0.00/0.00\\
        \rowcol T4-4 & 0.00/0.00 & {\bf 1.00}/{\bf 1.00} & \err/\err & 0.00/0.00\\
        T4-5 & 0.00/0.00 & {\bf 1.00}/{\bf 1.00} & \err/\err & 0.00/0.00\\
        \bottomrule
    \end{tabular}
    \vspace{-0.1cm}
\end{table} 

Tables~\ref{tab:trojanResults1000} and \ref{tab:trojanResults10000} shows the results of our experiments for trojan attacks. 
The tables again depict the probabilities with which the testing tool was or was not able to find a test input falsifying the property under interest. 
We considered two architectures for neural network models (NN1 and NN2), trained on the MNIST data set enhanced by 1,000 and 10,000 additional 
``poisened'' instances, respectively for the two tables, using 4 different trigger features T1 to T4 and 2 different target predictions (classes 4 and 5). 
The triggers are hence named T1-4, T1-5 and so on. 

It turned out that the property-based testing tool which we employed is not able to generate test cases at all. 
On all instances, it stopped with the error message ``hypothesis.errors.Unsatisfiable: Unable to satisfy assumptions of hypothesis'', typically after trying to generate test inputs for around 40 minutes. 
We suspect that the reason for this failure is the high number of features (100) in this data set, i.e., the fact that Hypothesis has 
to generate inputs for a function with 100 arguments which is likely not the setting envisaged by the developers of this tool. 

In order to be able to compare our technique to other methods, we hence decided to develop a prototype tool for 
adaptive random testing~\cite{DBLP:conf/asian/ChenLM04} with respect to trojan attacks.  
Note that there is no adaptive random tester allowing to specify arbitrary properties to be tested.  
Thus, a new implementation is required for every property, in particular the definition and implementation of a distance metric 
(for which we here used the euclidian distance metric on feature vectors). 
Tables~\ref{tab:trojanResults1000} and \ref{tab:trojanResults10000} therefore also give the result for our prototype adaptive random tester (ART). 
Interestingly, ART is able to find counterexamples for a number of models trained on the data set enhanced with 1,000 poisened instances, 
even more often than {\sc MLCheck} with a decision tree. 
However, in the harder cases with models trained on the data set enhanced with 10,000 instances, ART also produces no test inputs at all.

\llbox{In summary, {\sc MLCheck} outperforms other tools in almost all cases, even when they are specialised to the property to be tested.}

\pgfplotsset{
	compat=1.15,
	width=\columnwidth,
	height=6cm
}

\begin{figure}[t]
	\begin{center}
		\begin{tikzpicture}[scale=0.9]
             \definecolor{clr1}{rgb}{0.0, 0.5, 0.0}
             \definecolor{clr2}{rgb}{0.76, 0.13, 0.28}
             \definecolor{clr3}{rgb}{1.0, 0.49, 0.0}
		\definecolor{clr4}{rgb}{0.0, 0.0, 1.0}
		
		\begin{axis}[
		legend pos=north west, 
		ytick = {1,100,500,1000},
		ylabel={runtime (in seconds)},
		xtick = {1, 2, 3, 4, 5, 6, 7, 8, 9, 10, 11, 12}, 
		xmin=1, xmax=12,
		ymin = 1, ymax=1000,
		xticklabels = {1, 2, 3, 4, 5, 6, 7, 8, 9, 10, 11, 12}, 
		xlabel = {Solved tasks (ordered by runtime of approach)}
		]
		\addplot
		[clr1, line width=0.3mm, dotted] coordinates{
			(1, 1) (2, 8) (3, 8.43) (4, 242.99) (5, 246.78) (6, 258.26) (7, 301.32) (8, 348.98) (9, 367.03) (10, 383.58) (11, 390.32) (12, 657.78)
		};
		\addplot
		[clr2, line width=0.3mm, dashdotted] coordinates{
			(1, 8.97) (2, 30.54) (3, 42.17) (4, 73.89) (5, 76.88) (6, 80.78) (7, 141.79) (8, 282.56) (9, 287.98) (10, 387.93) (11, 562.62) (12, 653.61)
		};
		
		\addplot
		[clr3, line width=0.3mm] coordinates {
			(1, 92.57) (2, 125.75) (3, 126.27) (4, 159.53) (5, 172.31) (6, 180.26) (7, 195.16) (8, 199.45) (9, 207.94) (10, 212.58) 
		};
	    \addplot
	    [clr4, line width=0.3mm, loosely dashed] coordinates {
			(1, 4.39) (2, 5) (3, 13.67) (4, 19.53) (5, 93.56) (6, 120) (7, 179) (8, 548.98) (9, 982.33) (10, 1000.67)
	    };
		
		\legend{{\sc MLCheck\_dt}, {\sc MLCheck\_nn}, SG, AQ}
		\end{axis}
		\end{tikzpicture}
	\end{center}
	\caption{Runtime for checking fairness}
	\label{fig:execution-time-fair}
	\vspace{-0.1cm}
\end{figure}
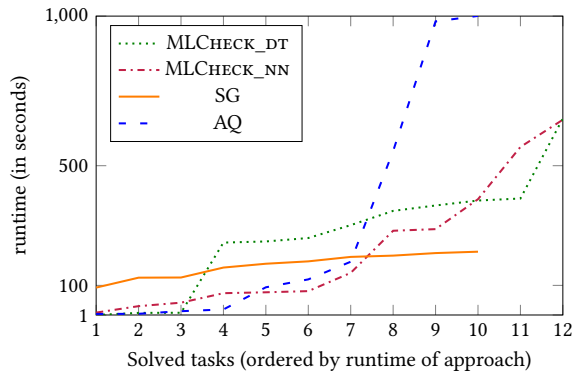

\begin{figure}[t]
	\begin{center}
		\begin{tikzpicture} [scale=0.9]
		\definecolor{clr1}{rgb}{0.0, 0.5, 0.0}
             \definecolor{clr2}{rgb}{0.76, 0.13, 0.28}
             \definecolor{clr3}{rgb}{1.0, 0.49, 0.0}
		\begin{axis}[
		legend pos=north west, 
		ytick = {1,100,500,1000},
		ylabel={runtime (in seconds)},
		xtick = {1, 6, 12, 18, 24, 30, 36}, 
		xmin=1, xmax=36,
		ymin = 1, ymax=1000,
		xticklabels = {1, 6, 12, 18, 24, 30, 36}, 
		xlabel = {Solved tasks (ordered by runtime of approach)}
		]
		\addplot
		[clr1, line width=0.3mm] coordinates{
			(1, 1) (2, 1.21) (3, 1.21) (4, 1.23) (5, 1.45) (6, 1.46) (7, 1.56) (8, 1.6) (9, 1.78) (10, 1.8) (11, 1.89) (12, 2.98) (13, 4.05) (14, 4.11) (15, 4.21) (16, 4.33) (17, 4.35) (18, 4.53) (19, 4.67) (20, 8.13) (21, 8.48) (22, 8.74) (23, 12.27) (24, 13.89) (25, 18.11) (26, 18.4) (27, 18.52) (28, 18.56) (29, 24.19) (30, 26.63) (31, 59.89) (32, 83.81) (33, 85.07) (34, 100.76) (35, 103.56) (36, 103.89)
		};
		\addplot
		[clr2, line width=0.3mm, dashdotted] coordinates{
			(1, 5.35) (2, 5.88) (3, 7.37) (4, 9.67) (5, 14.94) (6, 17.85) (7, 20.59) (8, 60.49) (9, 61) (10, 71.06) (11, 83.38) (12, 86.87) (13, 90.05) (14, 113.07) (15, 129.39) (16, 197.82) (17, 210.98) (18, 383.35) (19, 414.56) (20, 424.35) (21, 426.53) (22, 490.72) (23, 491.68) (24, 495.29) (25, 498.48) (26, 499.58) (27, 500.83) (28, 501.72) (29, 503.89) (30, 504.2) (31, 589.87) (32, 592.87) (33, 601.78) (34, 603.78) (35, 603.89) (36, 610.45)
		};
		
		\addplot
		[clr3, line width=0.3mm, loosely dashed] coordinates {
			(1, 13.66) (2, 18.69) (3, 19.22) (4, 26.55) (5, 26.74) (6, 27.61) (7, 29.39) (8, 29.55) (9, 32.79) (10, 188.03) (11, 191.25) (12, 192.03) (13, 219.51) (14, 260.41) (15, 260.97) (16, 261.39) (17, 262.68) (18, 263.28) (19, 267.72) (20, 334.47) (21, 374.47) (22, 375.55) (23, 379.06) (24, 382.73) (25, 385.04) (26, 428.04) (27, 451.52) (28, 671.35) (29, 817.22) (30, 913.56) (31, 967.89) (32, 982.98) (33, 987.67) (34, 988.87) (35, 988.88) (36, 989.12)
		};
		
		\legend{{\sc MLCheck\_dt}, {\sc MLCheck\_nn}, PBT}
		\end{axis}
		\end{tikzpicture}
	\end{center}
	\caption{Runtime for checking concept relationships}
	\label{fig:execution-time-cr}
	\vspace{-0.1cm}
\end{figure}
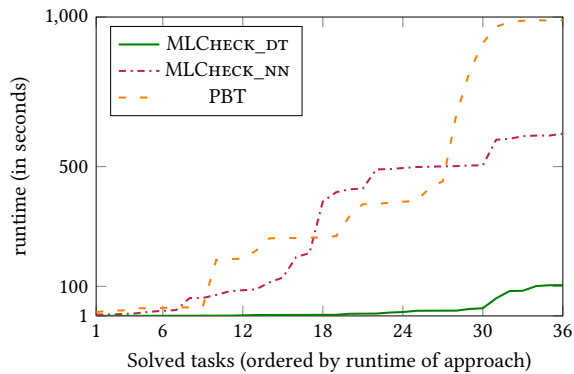

\smallskip
\noindent 
For {\bf RQ2}, we compared the efficiency (in terms of runtime) of the tools in generating test inputs violating the property under interest. 
We again report on the results separately for every application area 
and again the given values are averaged over 20 runs. 

Figure~\ref{fig:execution-time-fair} depicts the runtime of four tools on 12 fairness testing tasks (6 classifiers on the two datasets). 
SG and AEQUITAS curves end at 10 tasks as they do not run on models generated by one of the fair-aware algorithms. 
The x-axis depicts the number of tasks solved, ordered by runtime per tool from fastest to slowest, the y-axis is the runtime in seconds. 
We see that there are a number of tasks for which SG has a smaller runtime, though not significantly. 
SG uses SMT solving on decision trees as well, but does not consider the entire tree but only some paths. 
This explains the lower runtime for some tasks, but also the smaller numbers of generated failing test inputs. 
AEQUITAS has significantly larger runtimes than {\sc MLCheck} on the more complex tasks. 

Figure~\ref{fig:execution-time-cr} shows the runtime for testing concept relationships on 36 tasks 
(2 classifiers trained on 6 data sets and 3 properties to be checked). 
Here, the decision tree version of {\sc MLCheck} performs best and property-based testing 
is comparable to the neural network version for all but the  8 most complicated tasks. 

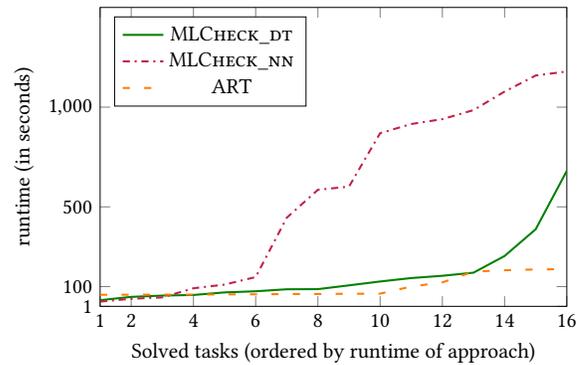
\begin{figure}[t]
    \begin{center}
        \begin{tikzpicture}[scale=0.9]
        \definecolor{clr1}{rgb}{0.0, 0.5, 0.0}
        \definecolor{clr2}{rgb}{0.76, 0.13, 0.28}
        \definecolor{clr3}{rgb}{1.0, 0.49, 0.0}
        \begin{axis}[
        legend pos=north west,
        ytick = {1,100,500,1000},
        ylabel={runtime (in seconds)},
        xtick = {1, 2, 4, 6, 8, 10, 12, 14, 16},
        xmin=1, xmax=16,
        ymin = 1, ymax=1500,
        xticklabels = {1, 2, 4, 6, 8, 10,  12, 14, 16},
        xlabel = {Solved tasks (ordered by runtime of approach)}
        ]
        \addplot
        [clr1, line width=0.3mm] coordinates{
            (1, 31.91) (2, 48.47) (3, 54.92) (4, 58.01) (5, 70.85) (6, 76.55) (7, 86.61) (8, 87.24) (9, 106.25) (10, 125.61) (11, 142.88) (12, 154.21) (13, 169.79) (14, 253.48) (15, 387.56) (16, 680.48)
        };
        \addplot
        [clr2, line width=0.3mm, dashdotted] coordinates{
            (1, 24.68) (2, 38.88) (3, 45.79) (4, 91.67) (5, 109.94) (6, 147.85) (7, 445.59) (8, 585.49) (9, 601.56) (10, 869.38) (11, 915.38) (12, 939.87) (13, 985.05) (14, 1078.07) (15, 1158.75) (16, 1178.82)
        };

        \addplot
        [clr3, line width=0.3mm, loosely dashed] coordinates {
            (1, 59.04) (2, 59.32) (3, 59.70) (4, 59.71) (5, 60.88) (6, 61.19) (7, 62.75) (8, 62.85) (9, 63.98) (10, 64.65) (11, 100.23) (12, 120.98) (13, 175.27) (14, 181.98) (15, 185.89) (16, 187.73)
        };

        \legend{{\sc MLCheck\_dt}, {\sc MLCheck\_nn}, ART}
        \end{axis}
        \end{tikzpicture}
    \end{center}
    \caption{Runtime for checking trojan attacks}
    \label{fig:execution-time-trojan}
    \vspace{-0.1cm}
\end{figure}

For trojan attacks, Figure~\ref{fig:execution-time-trojan} shows the runtimes of our two {\sc MLCheck} instances 
as well as our own adaptive random tester. 
It shows that the runtime of ART is relatively low compared to that of {\sc MLCheck} with a  neural network as white-box model. 
The reason for this is that our approach includes the training of the white-box model which for neural networks requires 
some time.

\llbox{Summarizing,  except for trojan attacks we see that the increased effectiveness of {\sc MLCheck} does not come at the prize of a much higher runtime. }

For {\bf RQ3}, we take another look at the tables of detected violations and figures of runtimes, now comparing the decision tree and neural 
network version of {\sc MLCheck}. 

With respect to the number of detected violations, we see that the neural network as white-box model is--with a few exceptions--only able to outperform the decision tree in case of the MUT being a neural network itself.
The better performance in these cases does often not come at the prize of  a (much) increased runtime. 
In some such cases the neural network white-box model can even spectacularly outperform the decision tree, 
which can be seen in the trojan attack setting when trained with 10,000 additional poisened instances. 
For the trojan attacks with a high number of features, the neural network white-box is much better in approximating the MUT. 
This confirmed our initial expectation that it does in fact make sense to employ two different white-box models 
with different generalization abilities  in test case generation. 

\smallskip
\noindent 
Note that all three application areas contain {\em hard} benchmarks characterised by only few counterexamples, 
either generated by using specific ML algorithms (fair-aware algorithms) or by assembling specific data sets,
obtained by embeddings (for concept relationships) or by flooding the training set with property-satisfying instances (trojan attacks). 
We can summarize the results of the experiments on these benchmarks as follows. 

\llbox{{\sc MLCheck} in particular outperforms other tools on hard testing tasks.}

\section{Discussion}

We briefly discuss some further aspects of our approach. 

\subsection{Soundness}

Our approach is sound in the sense of only generating test inputs which are counter examples to the property 
{\em on the black-box model}. We might generate candidate counter examples which are only valid 
counter examples on the approximating white-box model, but such counter examples do not get into 
the test suite. All candidate counter examples are checked on the model under test.

Furthermore, our approach allows to add {\em constraints} on feature values to the test input generation process. 
Such constraints can for instance reflect the ``data semantics'' (e.g., a feature `age' not allowing for values above 100), 
if such value restrictions are explicitly given. 
Currently, we derive constraints on minimal and maximal feature values from the training set and 
take these into account when the parameter \verb+bound_cex+ is set to \verb+True+. 

With the direct incorporation of such constraints in the SMT solving process, we can avoid costly transformations of 
counter examples which other approaches need to perform~\cite{zhang2020white}.

\subsection{Scalability} 

Though we have image recognition as one of our case studies,
we do not intend to claim that our approach will in general be applicable to image classifiers. 
Classifiers on images with several thousands of bits and hence several thousands of features  
will pose difficulties for the SMT solver. 
Note however that the number of hidden layers of the model under test is {\em not} a limiting factor 
as we employ a much simpler neural network for approximation, and only this needs translation to logical formulae.

Also note that our approach can so far not be applied to test for statistical properties of models and to  
models which operate on data {\em streams} (like audio processing). 

\subsection{Threats to Validity} 

The results discussed here depend on the chosen data sets, classifiers and properties. 
We have taken data sets and non-stochastic properties out of three areas as to demonstrate the tool's generality. 
The classifers are taken from standard ML libraries which will also be used by software developers for learning. 
We can however not exclude that we would get different findings on other areas using different training algorithms.  

The concrete numbers in the tables are affected by the {\em randomness} inside the ML algorithms, 
both for training MUTs and training our white-box models. To mitigate this thread, we ran every experiment 20 times and have given the mean together with the standard error of the mean.

%% file: related.tex
\section{Related work}\label{sec:related}

We briefly discuss other approaches to the validation of machine learning models.  

The most frequently studied type of models are deep neural networks (DNNs). 
In the area of testing neural networks one focus has been on the development of appropriate 
{\em coverage criteria} for DNNs, ranging from neuron coverage~\cite{PeiCYJ17} over 
some form of MC/DC coverage~\cite{SunWRHKK18} (in concolic testing) to 
multi-granularity testing criteria on test beds~\cite{DBLP:conf/kbse/MaJZSXLCSLLZW18}. 

The most frequently studied property of DNNs is 
{\em adversarial robustness}. It describes the vulnerability of a DNN to adversarial attacks, i.e.,
attacks in which a small deviation from a correctly classified input yields a different class prediction. 
Attack methods aim at generating such adversarial examples, and can be classified into 
white-box (e.g.,~\cite{DBLP:journals/corr/GoodfellowSS14,DBLP:conf/cvpr/Moosavi-Dezfooli16}) and 
black box  (e.g.,~\cite{deepsearch,DBLP:conf/ccs/PapernotMGJCS17}) approaches. 
Recent works have also proposed methods for computing probabilistic guarantees on robustness~\cite{DBLP:conf/aaai/CardelliKLP19}. 

Testing methods typically aim at generating counterexamples to properties. 
{\em Formal verification} on the other hand aims at correctness guarantees. 
For neural networks a number of verification techniques have been developed,
based on abstract interpretation~\cite{GehrMDTCV18,DBLP:journals/pacmpl/SinghGPV19,DBLP:conf/cav/ElboherGK20},
by layer-wise computations of safety constraints~\cite{HuangKWW17} or 
by a combination of SAT solving and linear programming~\cite{DBLP:conf/atva/Ehlers17}.

More recently, Pham et al.~\cite{DBLP:journals/corr/abs-2007-11206} have proposed a unifying framework to verify different types of neural network models (i.e.~models with different types of layers and activation functions) using two verification methods, namely {\em optimization-based falsification} and {\em statistical model checking}. Alike us, they also provide a specification language to specify several properties (such as fairness, robustness etc.) to be checked on the model.

Baluta et al.~\cite{DBLP:conf/ccs/BalutaSSMS19} propose {\em quantitative} verification for neural networks, i.e.,
verification which gives a quantitative account on the number of inputs violating some property. 
Similar to the encoding of   our NN white-box models, they translate  neural networks to logical formulae on which approximate model counting can then  provide 
estimates about the number of satisfying logical models. To make their approach scale, they apply it to 
binarized NNs only, and perform further quantization. 
Other verification approaches using logical encodings of DNNs together with SAT, SMT or MIP solvers   for property checking  
have been proposed by Narodytska et al.~\cite{DBLP:conf/aaai/NarodytskaKRSW18} (studying various properties, 
in particular also adversarial robustness), Pulina et al.~\cite{DBLP:conf/cav/PulinaT10} or Cheng et al.~\cite{DBLP:conf/atva/ChengNR17}. 
Katz et al.~\cite{DBLP:conf/cav/KatzBDJK17} in addition build a specific SMT solver for solving linear real arithmetic constraints arising from DNNs with ReLU activation functions. 
A study comparing different verification approaches of NNs and evaluating the impact of pruning techniques has been performed by Guidotti et al.~\cite{DBLP:journals/corr/abs-2003-07636};  
further surveys of validation approaches for neural networks have been done  by Liu et al.~\cite{DBLP:journals/corr/abs-1903-06758} and Huang et al.~\cite{survey20}. 
 
These approaches are specific to neural networks. 
A survey on testing techniques for ML models in general, i.e., not restricted to neural networks, has recently been assembled by Zhang et al.~\cite{DBLP:journals/corr/abs-1906-10742}.  
Model agnostic but property specific approaches most often target {\em fairness testing}, more precisely testing for individual discrimination. 
 Themis~\cite{galhotra2017fairness,DBLP:conf/sigsoft/AngellJBM18} is an automated test case generation technique that uses random testing complemented by three optimization procedures. It allows for checking two types of fairness definitions,  namely {\em causal} (i.e.~individual) discrimination and group discrimination for a given black-box model. Later, Aggarwal et al.~propose a symbolic approach (SG) to generate test cases for checking individual discrimination. They use a tool called LIME to generate a path of a decision tree from the MUT, on which they use dynamic symbolic execution to generate test cases.
 They show that their approach outperforms Themis and AEQUITAS~\cite{DBLP:conf/kbse/UdeshiAC18}. 
We, on the other hand, approximate the entire MUT by a white-box model (either a decision tree or a neural network) 
and then compute the test cases on this model. Moreover, our approach can be used for checking not only individual discrimination but also other types of fairness (e.g.~{\em fairness through awareness}~\cite{verma2018fairness}) as well as completely different properties.

%% file: conclusion.tex
\section{Conclusion}\label{sec:conclusion}

In this paper, we have proposed an approach and tool for property-driven testing of machine learning models. 
The approach encompasses a language for property specification and a method for targeted generation of test cases falsifying the property.
As future work, we intend to study the applicability of the more advanced verification techniques on deep neural networks (e.g.~\cite{DBLP:journals/corr/abs-2007-11206}) for 
generating counterexamples for our white-box model neural network. 
